 \documentclass[aps,pra,superscriptaddress,amsmath,amssymb,preprintnumbers,twocolumn,floatfix,showpacs,showkeys,10pt]{revtex4-1}
 \usepackage{amssymb} \usepackage{epsfig}
 \begin{document}
  \title{Pairwise quantum discord for a symmetric multi-qubit system  in different types of noisy channels  }

\author{You-neng Guo}
\email{guoxuyan2007@163.com}
\affiliation{ Department of Electronic and Communication Engineering, Changsha University, Changsha, Hunan
410003, People's Republic of China}
\author{Mao-fa Fang}
\email{mffang@hunnu.edu.cn}
\affiliation{ Key Laboratory of Low-Dimensional Quantum Structures and
Quantum Control of Ministry of Education, and Department of Physics,
Hunan Normal University, Changsha 410081, People's Republic of
China}
 \author{Ke Zeng}
  \affiliation{ Department of Electronic and Communication Engineering, Changsha University, Changsha, Hunan
410003, People's Republic of China}
\author{Guo-you Wang}
\affiliation{College of Science, Hunan University of Technology, Zhuzhou 412008, People's Republic of China}
 \begin{abstract}
We study the pairwise quantum discord (QD) for a symmetric multi-qubit system in  different types of noisy channels, such as phase-flip, amplitude damping, phase-damping, and depolarizing channels. Using the QD and geometric measure of quantum discord (GMQD) to quantify quantum correlations, some analytical or
numerical results are presented. The results show that, the dynamics of  the pairwise QD is related to the number of spin particles $N$ as well as initial parameter $\theta$ of the one-axis twisting collective state. With the number of spin particles $N$ increasing, the amount of the pairwise QD increases. However, when the amount of the pairwise QD arrives at a stable maximal value, the pairwise QD is independence of the number of spin particles $N$ increasing. The behavior of the pairwise QD is symmetrical during a period $0\leq \theta \leq 2\pi$. Moreover, we compare the pairwise QD dynamics with the GMQD for a symmetric multi-qubit system in different types of noisy channels.
 \end{abstract}

 \maketitle
\section{Introduction}
As we all know, quantum correlations, such as quantum entanglement, have been proposed as the key resource present in certain quantum communication tasks and quantum computational models [1-2]. Many different quantum correlation measures have been proposed to detect the nonclassical correlations beyond entanglement. Among them, quantum discord (QD) which has initially been introduced by Olliver and Zurek [3], captures more general quantum correlations, and later the studies of QD have been vigorously investigated both theoretically [4-11] and experimentally [12,13].
However, evaluation of QD requires a potentially complex optimization procedure in general and analytical results have been obtain only in few restricted cases of $2\bigotimes2$ and $2\bigotimes N$ systems [14-22]. In order to overcome the optimization procedure of QD, Dakic et al. [23] introduced another measure quantum correlations named geometric quantum discord (GMQD)for a more general bipartite quantum system.

On the other hand, decoherence is an inevitable phenomena owe to quantum system inevitable interacting with its surrounding environment, which
gives rise to degradation of quantum coherence. In other word, decoherence is a major hindrance to practice quantum information processing. Hence, it is very interesting and necessary to study the effect of decoherence on the evaluation of quantum correlations. Recently the researches on dynamics of pairwise entanglement have been
made great progress in various fields [24-32]. For example, Wang et al. showed spin squeezing implies pairwise entanglement for arbitrary symmetric multi-qubit states [28,29]. Yin et al. investigated the pairwise quantum correlations of symmetric multi-qubit states by using the
geometric measure of quantum discord [30]. Xi et al. investigated pairwise quantum correlations for superpositions of Dicke states [32]. But pairwise correlations, as we know, quantified by QD and its
geometric measure, in a symmetric multi-qubit system under decoherence noisy channels seem to have been seldom
exploited before. It is of interest and importance to investigate how decoherence affects the pairwise correlations under decoherence noisy channels. Motivated from the recent study on the dynamics of QD and GMQD under the influence of external environments for different multipartite states [33], such as Werner-GHZ type three-qubit and six-qubit states, we here devote to examining the pairwise quantum correlation properties in terms of QD for pairs of particles extracted from a symmetric state. The two-particle density matrix is expressed in terms of expectation values of collective spin operators $S$ for the large system [28-30,34-40].

Our attention is mainly focused on  what happens to the pairwise QD for a symmetric multi-qubit system in four different types of noisy channels, such as phase-flip, amplitude damping, phase-damping, and depolarizing channels. We extend the results obtained in Ref. [40], where using the spin squeezing to witness multipartite correlations under decoherence is investigated. Using QD and GMQD to quantify quantum correlations, some analytical or
numerical results are presented. The results show that, the dynamics of  the pairwise QD is related to the number of spin particles $N$ as well as initial parameter $\theta$ of the one-axis twisting collective state. With the number of spin particles $N$ increasing, the amount of the pairwise QD increases. However, when the amount of the pairwise QD arrives at a stable maximal value, it is independence of the number of spin particles $N$ increasing. The behavior of the pairwise QD is symmetrical during a period $2\pi$. Moreover, we compare the pairwise QD dynamics with the geometric measure of GMQD for a symmetric multi-qubit system in different types of noisy channels.

The layout is as follows: In Sec. \textrm{II}, we briefly review the definition of QD and GMQD, and give the expression of QD and GMQD. We illustrate initial states and noise channels in  Sec. \textrm{III}. We devote to examining their pairwise quantum correlation properties in terms of QD for pairs of particles extracted from a symmetric state of multi-qubit systems in different types of noisy channels in Sec. \textrm{IV}. Finally,
we give the conclusion  in Sec. \textrm{V}.
 \section{QD and GMQD expression of two qubits }  
Before starting to discuss the dynamics of quantum correlations, we review the correlation measures used in our investigation, namely, QD and GMQD.
Let us start with the definition of QD for any bipartite quantum X-state $\rho_{AB}$
\begin{equation}\rho_{AB}= \left(
\begin{array}{ c c c c l r }
\rho_{11} & 0 & 0 & \rho_{14} \\
0 & \rho_{22} & \rho_{23} & 0 \\
0 & \rho_{32} & \rho_{33} & 0 \\
\rho_{41} & 0 & 0 & \rho_{44} \\
\end{array}
\right),
\end{equation}
then, the classical correlation $C(\rho_{AB})$ is defined as
\begin{equation}
C(\rho_{AB})=S(\rho_{A})-\min_{\{\prod_{k}^{B}\}}S(\rho_{A|B}),
\end{equation}
where $S(\rho)=-tr(\rho \log_{2}\rho)$ is the von Neumann entropy. Moreover, the minimum is taken over the set of positive operator valued measurement $\{\prod_{k}^{B}\}$
on subsystem B,  $S(\rho_{A|B})$ is the conditional entropy for the subsystem A.  QD can be simply obtained
by subtracting $C(\rho_{AB})$ from the total amount of correlation [3]
\begin{equation}
QD(\rho_{AB})=I(\rho_{A:B})-C(\rho_{AB}),
\end{equation}
where the total correlation is quantified by the quantum mutual information $I(\rho_{A:B})$
\begin{equation}
I(\rho_{A:B})=S(\rho_{A})+S(\rho_{B})+\Sigma_{i=1}^{4}\epsilon_{i}\log_{2}\epsilon_{i},
\end{equation}
where $\rho_{A}$($\rho_{B}$) is the reduced matrix of $\rho_{AB}$ by tracing out $B$($A$), and
\begin{eqnarray}
S(\rho_{A})&=&-(\rho_{11}+\rho_{22})\log_{2}(\rho_{11}+\rho_{22}) \nonumber\\
&-&(\rho_{33}+\rho_{44})\log_{2}(\rho_{33}+\rho_{44}),
\end{eqnarray}
\begin{eqnarray}
S(\rho_{B})&=&-(\rho_{11}+\rho_{33})\log_{2}(\rho_{11}+\rho_{33}) \nonumber\\
&-&(\rho_{22}+\rho_{44})\log_{2}(\rho_{22}+\rho_{44}).
\end{eqnarray}
The eigenvalues of the density matrix $\rho_{AB}$ in Eq. (11) are given by
\begin{equation}
\begin{array}{ c c c c l r }
\epsilon_{1}=\frac{1}{2}[(\rho_{11}+\rho_{44})+\sqrt{(\rho_{11}-\rho_{44})^2+4|\rho_{14}|^2}], &\\
\epsilon_{2}=\frac{1}{2}[(\rho_{11}+\rho_{44})-\sqrt{(\rho_{11}-\rho_{44})^2+4|\rho_{14}|^2}], &\\
\epsilon_{3}=\frac{1}{2}[(\rho_{22}+\rho_{33})+\sqrt{(\rho_{22}-\rho_{33})^2+4|\rho_{23}|^2}], &\\
\epsilon_{4}=\frac{1}{2}[(\rho_{22}+\rho_{33})-\sqrt{(\rho_{22}-\rho_{33})^2+4|\rho_{23}|^2}]. &\\
\end{array}
\end{equation}

Since the calculation of classical correlation involves a potentially complex optimization process, there exists no general analytical expression of discord even for the simplest case of two-qubit state. It is difficult to calculate the QD by analytical solutions, while for a bipartite quantum X-state
described by the density matrix $\rho_{AB}$,
we can derive an expression of the QD [22]
\begin{equation}
\begin{array}{ c c c c l r }
QD(\rho_{AB})=\min(Q_{1},Q_{2}),\\
\end{array}
\end{equation}

where

$Q_{i}=H(\rho_{11}+\rho_{33})+\Sigma_{i=1}^{4}\epsilon_{i}\log_{2}\epsilon_{i}+D_{j}$,

$D_{1}=H(\frac{1+\sqrt{[1-2(\rho_{33}+\rho_{44})]^2+4(|\rho_{14}|+|\rho_{23}|)^2}}{2})$,

$D_{2}=-\Sigma_{i}\rho_{ii}\log_{2}\rho_{ii}-H(\rho_{11}+\rho_{33})$,

$H(x)=-x\log_{2}x-(1-x)\log_{2}(1-x)$.

Besides, GMQD for any bipartite quantum X-state $\rho_{AB}$ has been proposed by Dakic et al. [23]. A bipartite quantum X-state $\rho_{AB}$ writes in Bloch representation as
\begin{eqnarray}
\rho_{AB}&=&\frac{1}{4} [\sigma_{0}\otimes \sigma_{0}+\sum_{i}^{3}(x_{i}\sigma_{i}\otimes \sigma_{0}+y_{i}\sigma_{0}\otimes \sigma_{i}) \nonumber\\
&+&\sum_{i,j=1}^{3}R_{ij}\sigma_{i}\otimes \sigma_{j} ],
\end{eqnarray}
where $x_{i}=Tr(\rho\sigma_{i}\otimes \sigma_{0})$, $y_{i}=Tr(\rho\sigma_{0}\otimes \sigma_{i})$ are components of local Bloch vectors and $R_{ij}=Tr(\rho\sigma_{i}\otimes \sigma_{j})$ are components of the correlation tensor. The operators $\sigma_{i} (i=1,2,3)$ stand for the three Pauli matrices and $\sigma_{0}$ is the identity matrix. The expression of the GMQD is given by
\begin{equation}
GMQD(\rho_{AB})=\frac{1}{4}(||x||^2+||R||^2-k_{max}),
\end{equation}
where $x=(x_{1},x_{2},x_{3})^T$, $R$ is the matrix with elements $R_{ij}$, and $k_{max}$ is the largest eigenvalues of matrix defined by
\begin{equation}
K=xx^{T}+RR^{T}.
\end{equation}
Note that the function $GMQD(\rho_{AB})$ reaches its maximal value of $1/2$ , the normalized GMQD takes the form $2GMQD(\rho_{AB})$ in this paper. In the following, we use Eq. (8) and Eq. (10) to discuss the pairwise quantum correlation dynamics under different types of noisy channels.
\section{Initial states and noise model}  
We consider a system of $N$ exchange symmetry spin-1/2 particles with the ground state $|0\rangle$ and excited $|1\rangle$. This system's properties can be described by the collective operators $S_{\alpha}=\sum_{i=1}^{N}S_{i\alpha}=\frac{1}{2}\sum_{i=1}^{N}\sigma_{i\alpha}$, for $\alpha=x,y,z$. The one-axis twisting Hamiltonian reads[40]

\begin{equation}
H=\chi S_{x}^{2},
\end{equation}
where $\chi$ is a nonlinear coupling constant for all pairwise particles interaction. We choose the initial symmetry state, which has been prepared in the product state $|0\rangle_{N}=|00...0\rangle$, based on its dynamic evolution, the one-axis twisting collective state with even parity at time $t$ is obtained as
\begin{equation}
|\psi(t)\rangle=\exp(-i\theta S_{x}^{2})|0\rangle_{N},
\end{equation}

where $\theta=\chi t$ is the one-axis twisting angle. Following the previous discussion, the states with exchange symmetry and parity ensures that its two-qubit reduced state can be extracted randomly from this state, in the basis of ${|00\rangle,|01\rangle,|10\rangle,|11\rangle}$, whose two-quibt reduced density matrix can be written form Eq. (1). The elements of the two-quibt reduced density matrix can be represented by the local expectation values for the one-axis twisting state [28-30,34-40]
\begin{equation}
\rho_{11}=\frac{1}{4}(1+2\langle\sigma_{1z}\rangle +\langle\sigma_{1z}\sigma_{2z}\rangle), \nonumber
\end{equation}
\begin{equation}
\rho_{22}=\rho_{33}=\frac{1}{4}(1-\langle\sigma_{1z}\sigma_{2z}\rangle), \nonumber
\end{equation}
\begin{equation}
\rho_{44}=\frac{1}{4}(1-2\langle\sigma_{1z}\rangle +\langle\sigma_{1z}\sigma_{2z}\rangle),\nonumber
\end{equation}
\begin{equation}
\rho_{23}=\rho_{32}^{*}=\langle\sigma_{1+}\sigma_{2-}\rangle, \nonumber
\end{equation}
\begin{equation}
\rho_{14}=\rho_{41}^{*}=\langle\sigma_{1-}\sigma_{2-}\rangle.
\end{equation}
where the local expectation values for the one-axis twisting state were given by

\begin{equation}
\langle\sigma_{1z}\rangle=-\cos^{N-1}(\frac{\theta}{2}),\nonumber
\end{equation}
\begin{equation}
\langle\sigma_{1z}\sigma_{2z}\rangle=\frac{1}{2}(1+\cos^{N-2}\theta), \nonumber
\end{equation}
\begin{equation}
\langle\sigma_{1+}\sigma_{2-}\rangle=\frac{1}{8}(1-\cos^{N-2}\theta), \nonumber
\end{equation}
\begin{equation}
\langle\sigma_{1-}\sigma_{2-}\rangle=-\frac{1}{8}(1-\cos^{N-2}\theta)-\frac{i}{2}\sin(\frac{\theta}{2})\cos^{N-2}(\frac{\theta}{2}).
\end{equation}
Following, we will focus on decoherence of the one-axis twisting state when each spin particle is independently coupled to local noisy environment. Decoherence is a major hindrance to practice quantum information processing, it is important to discuss how decoherence affect the quantum correlation.
Here, we investigate what happens to the pairwise QD for a symmetric multi-qubit system in four different types of noisy channels, such as phase-flip, amplitude damping, phase-damping, and depolarizing channels. The dynamics of qubits interacting independently with individual environments is described by the solutions of the appropriate Born-Markov-Lindblad equations, which can also be obtained by so called the Kraus operator approach. Given the initial state $\rho$,
the state under noise is given by
\begin{equation}
\varepsilon(\rho)=\sum_{E1,E2,..EN}(\bigotimes_{i=1}^{N}K_{Ei})\rho(\bigotimes_{i=1}^{N}K_{Ei}^{\dagger})
\end{equation}
where $K_{Ei}$ denotes the Kraus operator for the $i-$th particle. Since the system has exchange symmetry, and the decoherence channels act independently on each particle. Therefore, all the elements of the two-quibt reduced density matrix are determined by some correlation functions and expectations, and any expectation value of the operator $A$ can be calculated as $\langle A \rangle = Tr[A \varepsilon(\rho)]=Tr [\varepsilon_{}^{\dagger}(\rho)A]$.

\subsection{Phase flip channel}

The Kraus operators for single qubit in phase-flip channel are given by
\begin{equation}E_{0}= \left(
\begin{array}{ c c c c l r }
\sqrt{1-p} & 0  \\
0 & \sqrt{1-p}  \\
\end{array}
\right),
\end{equation}
\begin{equation}E_{1}= \left(
\begin{array}{ c c c c l r }
\sqrt{p} & 0  \\
0 & -\sqrt{p}  \\
\end{array}
\right),
\end{equation}
where $p=\exp(-\gamma t)$.
Since the system has exchange symmetry, and the decoherence channels act independently on each particle. Therefore, substituting Eqs. (17) and (18) into Eq. (16), and combining Eq. (14) and (15), the elements of the density matrix can be represented by the local expectation values
\begin{equation}
\rho_{11}(t)=\frac{1}{4}(1+2\langle\sigma_{1z}\rangle +\langle\sigma_{1z}\sigma_{2z}\rangle),\nonumber
\end{equation}
\begin{equation}
\rho_{22}(t)=\rho_{33}=\frac{1}{4}(1-\langle\sigma_{1z}\sigma_{2z}\rangle), \nonumber
\end{equation}
\begin{equation}
\rho_{44}(t)=\frac{1}{4}(1-2\langle\sigma_{1z}\rangle +\langle\sigma_{1z}\sigma_{2z}\rangle), \nonumber
\end{equation}
\begin{equation}
\rho_{23}(t)=\rho_{32}^{*}(t)=(1-2p)^2\langle\sigma_{1+}\sigma_{2-}\rangle, \nonumber
\end{equation}
\begin{equation}
\rho_{14}(t)=\rho_{41}^{*}(t)=(1-2p)^2\langle\sigma_{1-}\sigma_{2-}\rangle.
\end{equation}
\subsection{Amplitude damping channel}
Amplitude damping channel which is used to characterize spontaneous emission describes the energy dissipation from a quantum system. The Kraus operators for a single qubit are given by
\begin{equation}E_{0}= \left(
\begin{array}{ c c c c l r }
\sqrt{p} & 0  \\
0 & 1  \\
\end{array}
\right),
\end{equation}
\begin{equation}E_{1}= \left(
\begin{array}{ c c c c l r }
0 & 0  \\
\sqrt{1-p} & 0  \\
\end{array}
\right),
\end{equation}
where $p=\exp(-\gamma t)$. According to previous analysis, the elements of the density matrix can be represented by the local expectation values
\begin{eqnarray}
\rho_{11}(t)&=&\frac{1}{4}[1+2(p\langle\sigma_{1z}\rangle +p-1) +(p^2\langle\sigma_{1z}\sigma_{2z}\rangle \nonumber\\
&-&2(1-p)p\langle\sigma_{1z}\rangle+(1-p)^2)], \nonumber
\end{eqnarray}
\begin{eqnarray}
\rho_{22}(t)=\rho_{33}&=&\frac{1}{4}[1-(p^2\langle\sigma_{1z}\sigma_{2z}\rangle-2(1-p)p\langle\sigma_{1z}\rangle\nonumber\\
&+&(1-p)^2)], \nonumber
\end{eqnarray}
\begin{eqnarray}
\rho_{44}(t)&=&\frac{1}{4}[1-2(p\langle\sigma_{1z}\rangle +p-1) +(p^2\langle\sigma_{1z}\sigma_{2z}\rangle \nonumber\\
&-&2(1-p)p\langle\sigma_{1z}\rangle+(1-p)^2)], \nonumber
\end{eqnarray}
\begin{eqnarray}
\rho_{23}(t)=\rho_{32}^{*}(t)=p\langle\sigma_{1+}\sigma_{2-}\rangle, \nonumber
\end{eqnarray}
\begin{eqnarray}
\rho_{14}(t)=\rho_{41}^{*}(t)=p\langle\sigma_{1-}\sigma_{2-}\rangle.
\end{eqnarray}
\subsection{Phase-damping channel}
Phase-damping channel describes a quantum noise with loss of quantum phase information without loss of energy. The Kraus operators for a single qubit are given by
\begin{equation}E_{0}= \left(
\begin{array}{ c c c c l r }
\sqrt{p} & 0  \\
0 & \sqrt{p}  \\
\end{array}
\right),
\end{equation}
\begin{equation}E_{1}= \left(
\begin{array}{ c c c c l r }
\sqrt{1-p}& 0  \\
0 & 0  \\
\end{array}
\right),
\end{equation}
\begin{equation}E_{2}= \left(
\begin{array}{ c c c c l r }
0& 0  \\
0 & \sqrt{1-p}  \\
\end{array}
\right),
\end{equation}
where $p=\exp(-\gamma t)$. According to previous analysis, the elements of the density matrix can be represented by the local expectation values
\begin{equation}
\rho_{11}(t)=\frac{1}{4}(1+2\langle\sigma_{1z}\rangle +\langle\sigma_{1z}\sigma_{2z}\rangle),\nonumber
\end{equation}
\begin{equation}
\rho_{22}(t)=\rho_{33}=\frac{1}{4}(1-\langle\sigma_{1z}\sigma_{2z}\rangle),\nonumber
\end{equation}
\begin{equation}
\rho_{44}(t)=\frac{1}{4}(1-2\langle\sigma_{1z}\rangle +\langle\sigma_{1z}\sigma_{2z}\rangle),\nonumber
\end{equation}
\begin{equation}
\rho_{23}(t)=\rho_{32}^{*}(t)=p^2\langle\sigma_{1+}\sigma_{2-}\rangle,\nonumber
\end{equation}
\begin{equation}
\rho_{14}(t)=\rho_{41}^{*}(t)=p^2\langle\sigma_{1-}\sigma_{2-}\rangle.
\end{equation}

\subsection{Depolarizing  channel}
Depolarizing  channel is another important type of quantum noise, it describes the process in which the density matrix is dynamically replaced by the state $I/2$. $I$ denoting identity matrix of a qubit. The Kraus operators for a single qubit are given by
\begin{equation}E_{0}=\sqrt{1-\frac{3p}{4}} \left(
\begin{array}{ c c c c l r }
1 & 0  \\
0 &1  \\
\end{array}
\right),
\end{equation}
\begin{equation}E_{1}= \sqrt{\frac{p}{4}}\left(
\begin{array}{ c c c c l r }
0& 1  \\
1 & 0  \\
\end{array}
\right),
\end{equation}
\begin{equation}E_{2}= \sqrt{\frac{p}{4}}\left(
\begin{array}{ c c c c l r }
0& -i  \\
i & 0  \\
\end{array}
\right),
\end{equation}
\begin{equation}E_{3}=\sqrt{\frac{p}{4}} \left(
\begin{array}{ c c c c l r }
1 & 0  \\
0 &-1  \\
\end{array}
\right),
\end{equation}
where $p=\exp(-\gamma t)$. According to previous analysis, the elements of the density matrix can be represented by the local expectation values
\begin{eqnarray}
\rho_{11}(t)&=&\frac{1}{4}[1+2((1-p)\langle\sigma_{1z}\rangle +\frac{p}{2}) +(1-p)^2\langle\sigma_{1z}\sigma_{2z}\rangle \nonumber\\ &+&(1-p)p\langle\sigma_{1z}\rangle+\frac{p^2}{4}],\nonumber
\end{eqnarray}
\begin{eqnarray}
\rho_{44}(t)&=&\frac{1}{4}[1-2((1-p)\langle\sigma_{1z}\rangle +\frac{p}{2}) +(1-p)^2\langle\sigma_{1z}\sigma_{2z}\rangle \nonumber\\ &+&(1-p)p\langle\sigma_{1z}\rangle+\frac{p^2}{4}],\nonumber
\end{eqnarray}
\begin{eqnarray}
\rho_{22}(t)=\rho_{33}&=&\frac{1}{4}[1-((1-p)^2\langle\sigma_{1z}\sigma_{2z}\rangle+(1-p)p\langle\sigma_{1z}\rangle\nonumber\\
&+&\frac{p^2}{4})],\nonumber
\end{eqnarray}
\begin{eqnarray}
\rho_{23}(t)=\rho_{32}^{*}(t)=(1-p)^2\langle\sigma_{1+}\sigma_{2-}\rangle,\nonumber
\end{eqnarray}
\begin{eqnarray}
\rho_{14}(t)=\rho_{41}^{*}(t)=(1-p)^2\langle\sigma_{1-}\sigma_{2-}\rangle.
\end{eqnarray}

\section{Discussion}  
In this section, we devote to examining their pairwise quantum correlation properties in terms of QD for pairs of particles extracted from a symmetric state of multi-qubit systems in different types of noisy channels, such as phase flip, amplitude damping, Phase-damping, and depolarizing channel, the two-particle density matrix is expressed in terms of expectation values of collective spin operators $S$ for the large system. Combining  Eq. (18)- Eq. (20), we can compute QD.

In Figure 1, we plot the pairwise QD of the one-axis twisting collective spin state given by Eq. (11) as a functions of $\gamma t$ and $N$ for fixed $\theta=0.1\pi$ in different types of noisy channels.  It can be seen that in phase-flip channel, the dynamics of pairwise QD decays quickly to zero and revives increasingly to its initial values. This result agrees with previous study of similar nature for GHZ states [33]. The revival of pairwise QD dynamics reaches to its initial values and keeps this stable values in phase-flip decoherence channel. This indicates the pairwise QD is not effected by phase-flip decoherence channel after The revival of pairwise QD dynamics reaches to its initial values. Besides, with the number of spin particles $N$ increasing, the amount of the pairwise QD increases. However, when the amount of the pairwise QD arrives at a stable maximal value, it is independence of the number of spin particles $N$ increasing. Compared to phase-flip channel, the pairwise QD dynamics decay in a monotonic fashion in amplitude damping, phase-damping, and depolarizing channel. This indicates that the amplitude damping, phase-damping, and depolarizing channel are heavily influence the pairwise QD. What is more, the pairwise QD dynamics decays more slowly in amplitude damping channel than those in Phase-damping, and depolarizing channel.

\begin{figure}[htpb]
  \begin{center}
  \includegraphics[width=7.5cm]{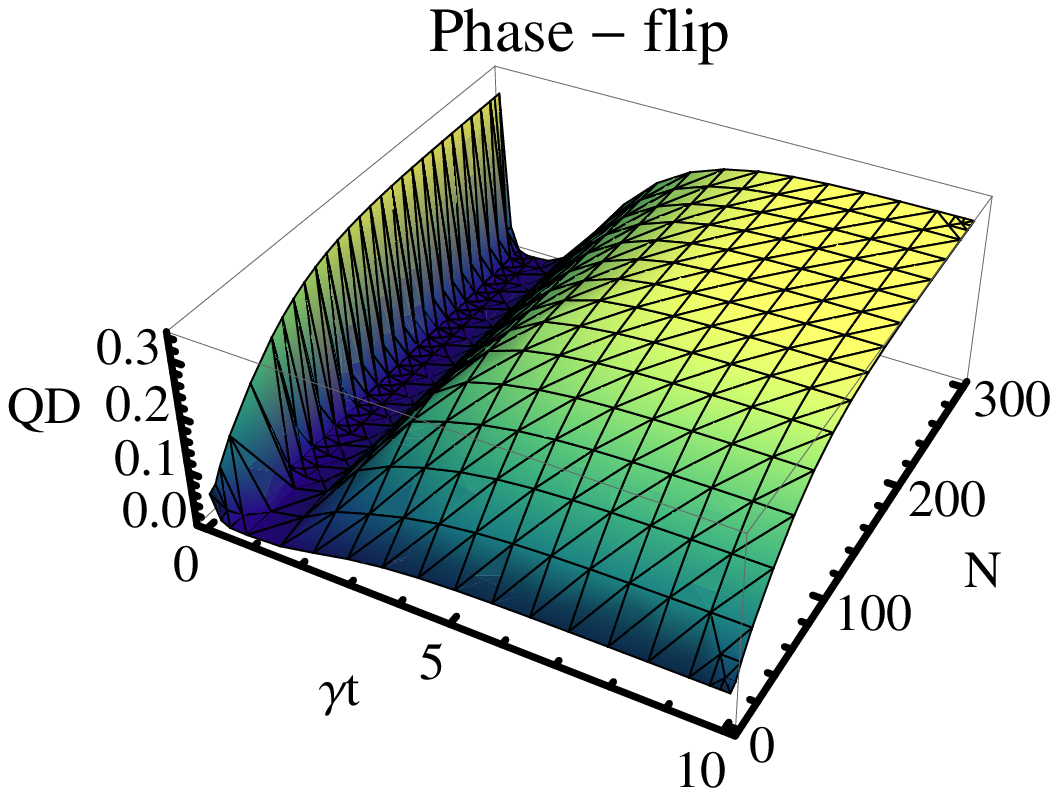}
     \includegraphics[width=7.5cm]{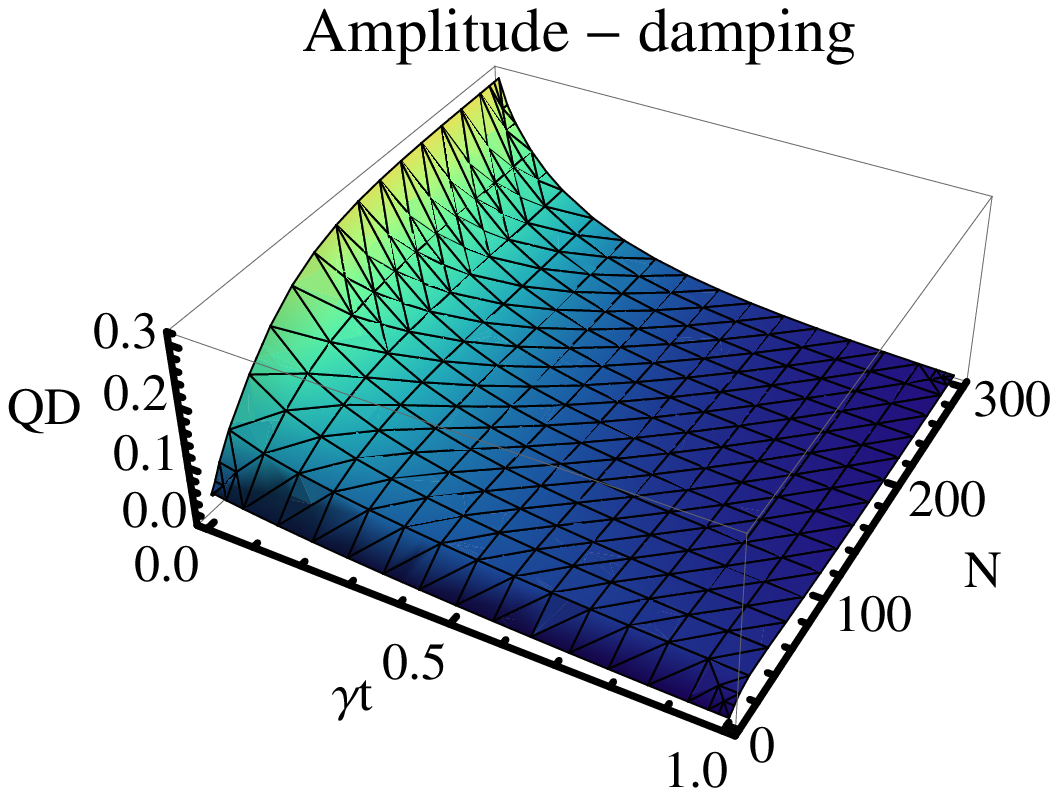}
     \includegraphics[width=7.5cm]{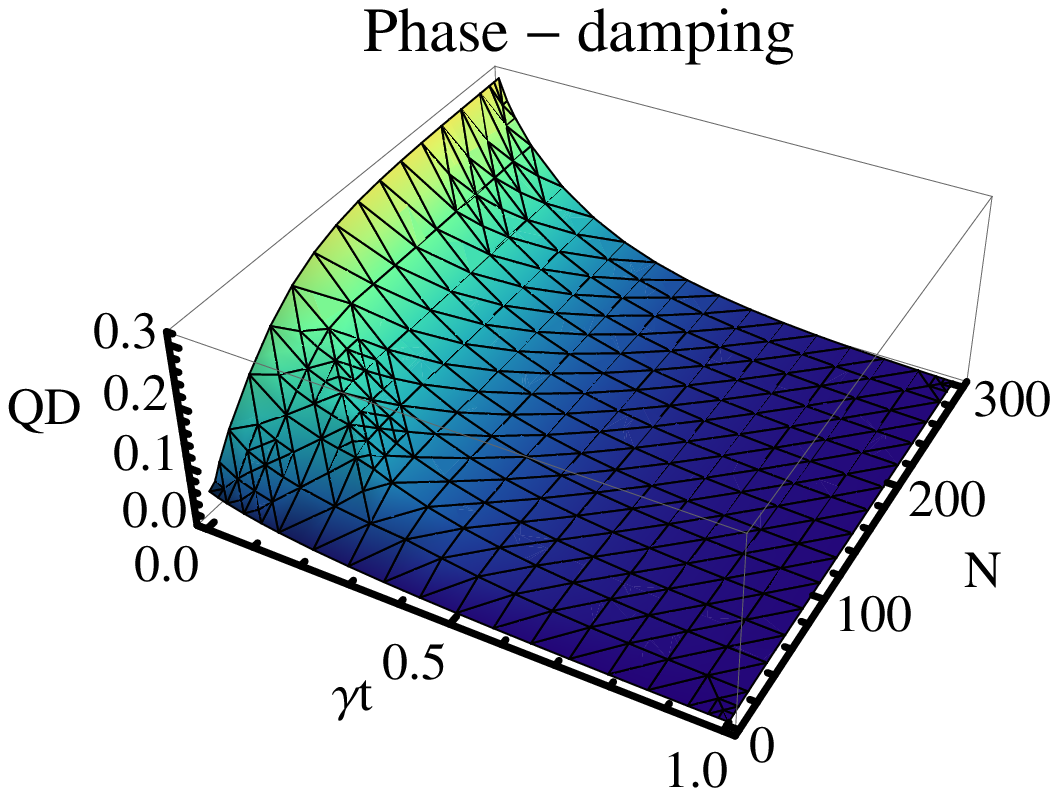}
     \includegraphics[width=7.5cm]{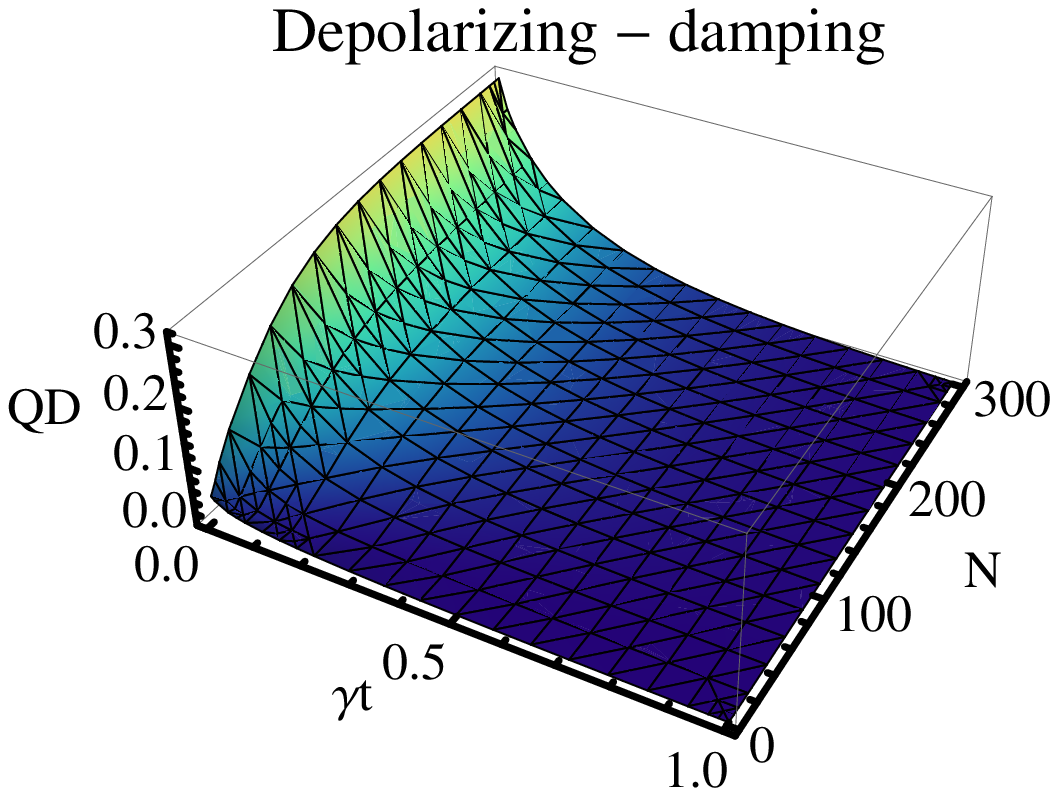}\quad
 \caption{\label{Fig1}(Color online)Quantum discord (QD) of the one-axis twisting collective spin state given by Eq. (11) as a functions of $\gamma t$ and $N$ for fixed $\theta=0.1\pi$ in different types of noisy channels.}
  \end{center}
\end{figure}

\begin{figure}[htpb]
  \begin{center}
   \includegraphics[width=7.5cm]{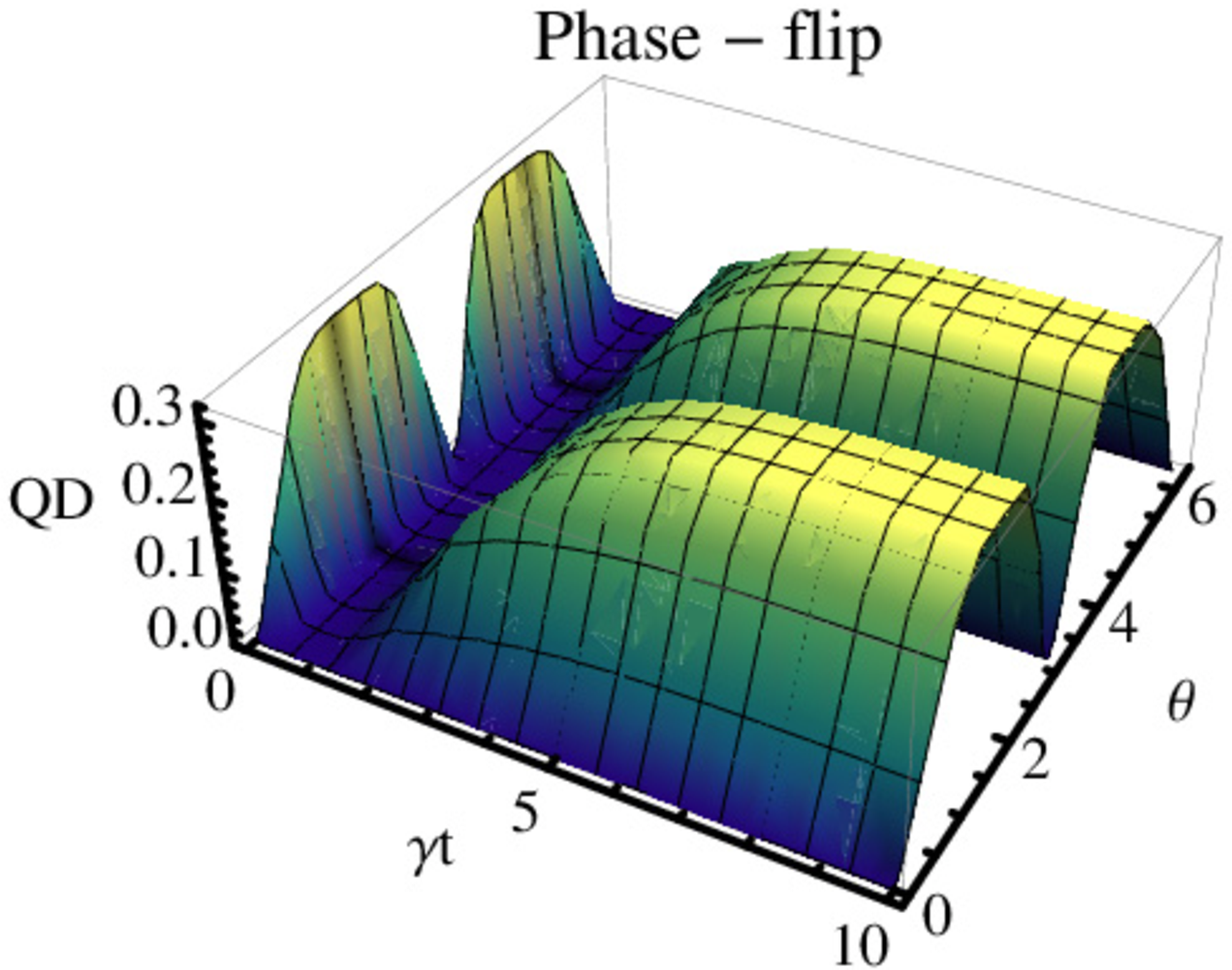}
     \includegraphics[width=7.5cm]{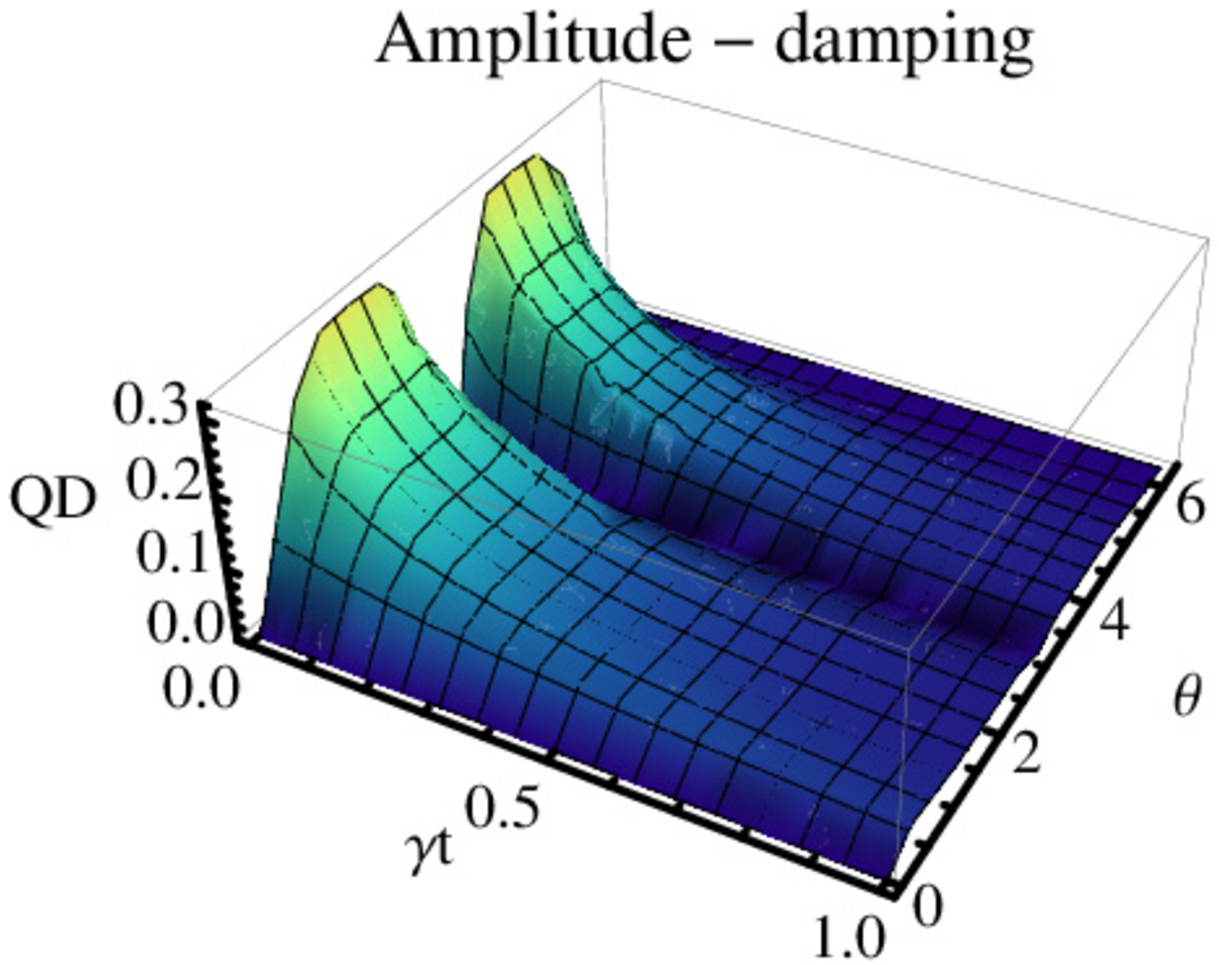}
     \includegraphics[width=7.5cm]{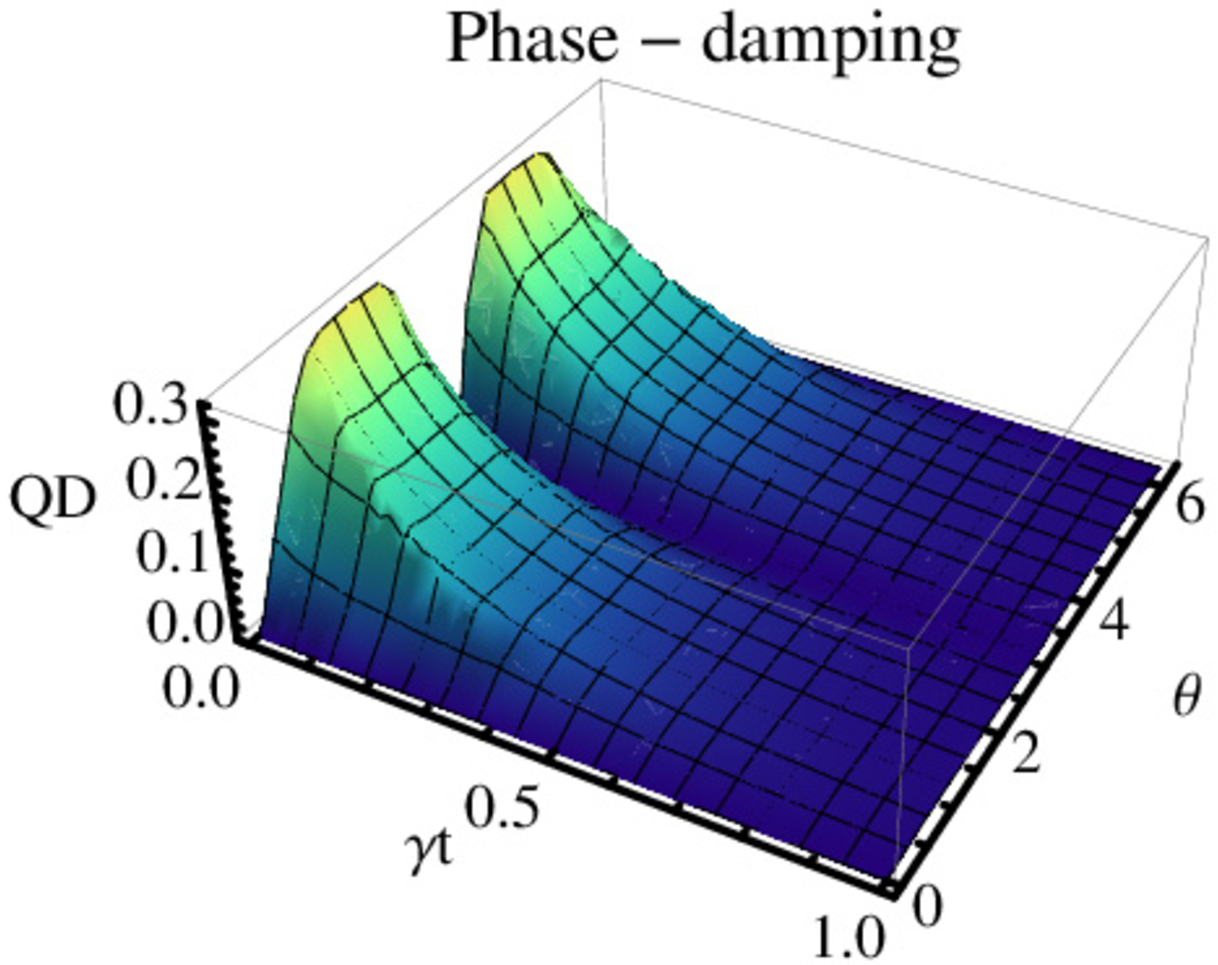}
    \includegraphics[width=7.5cm]{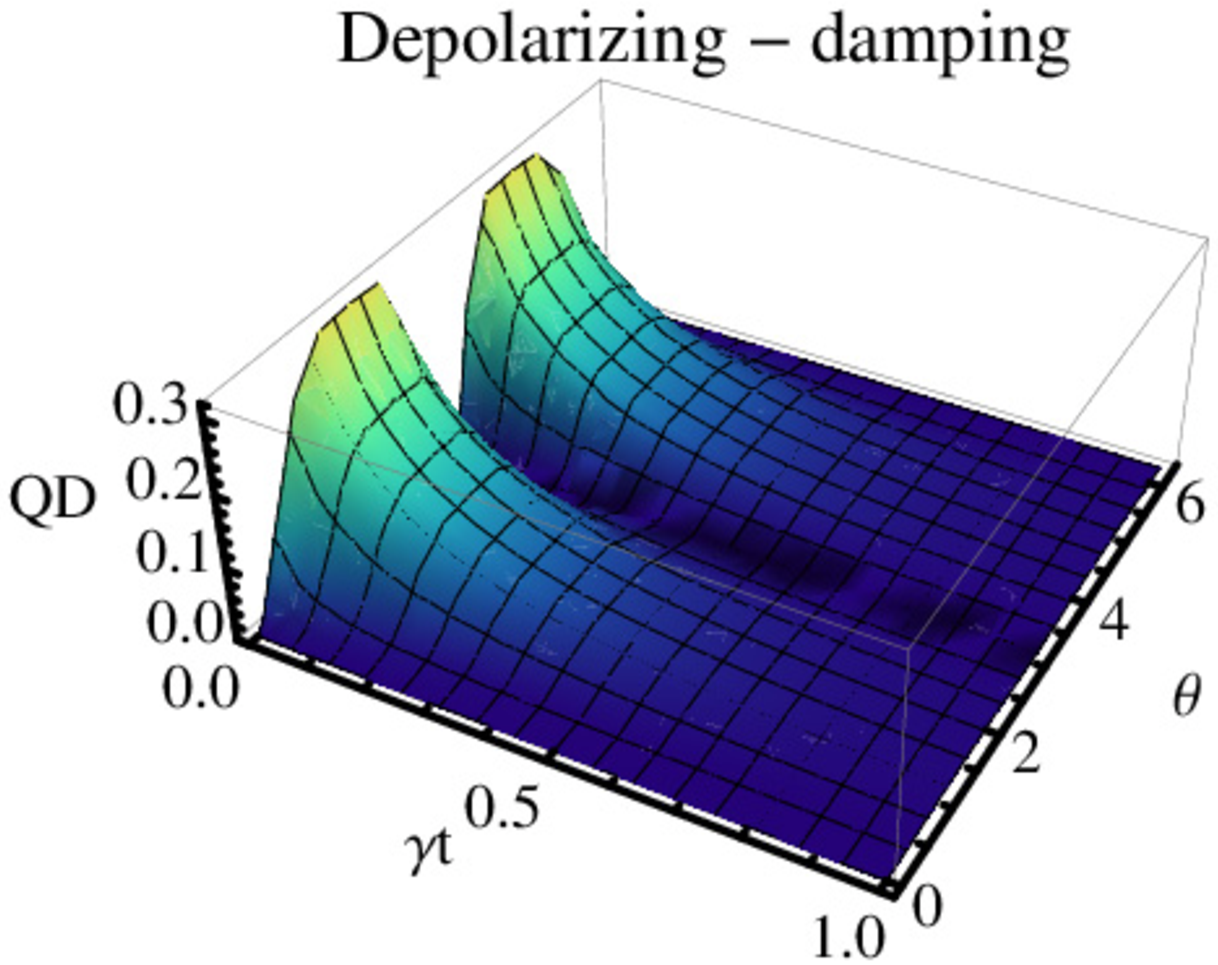}\quad
 \caption{\label{Fig2}(Color online)Quantum discord (QD) of the one-axis twisting collective spin state given by Eq. (11) as a functions of $\gamma t$ and $\theta$ for fixed $N=12$ in different types of noisy channels.}
\end{center}
\end{figure}
\begin{figure}[htpb]
  \begin{center}
 \includegraphics[width=7.5cm]{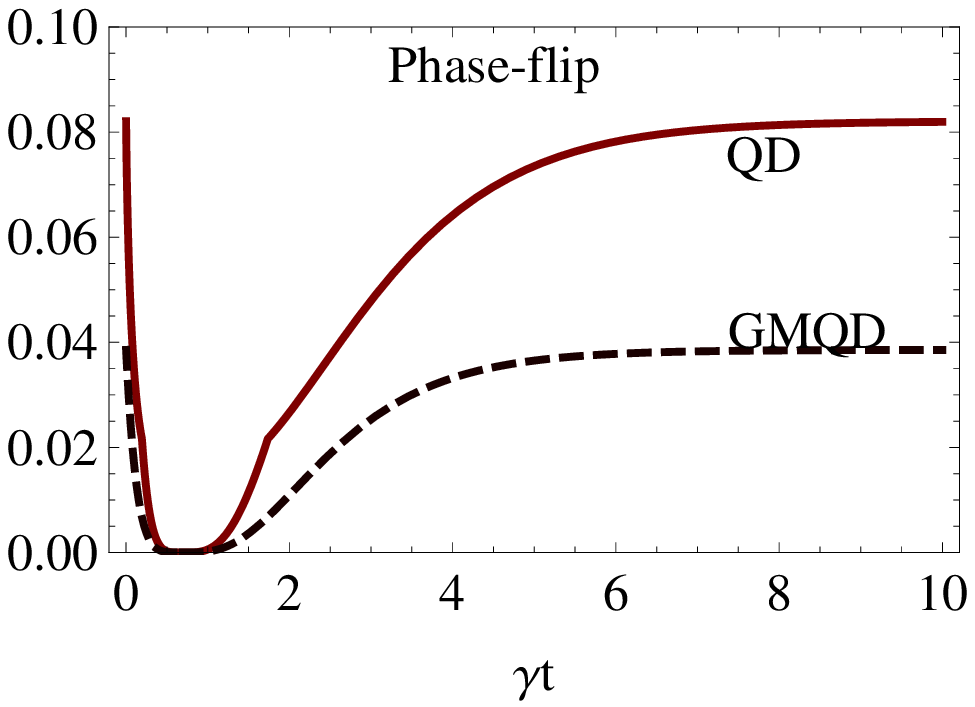}
     \includegraphics[width=7.5cm]{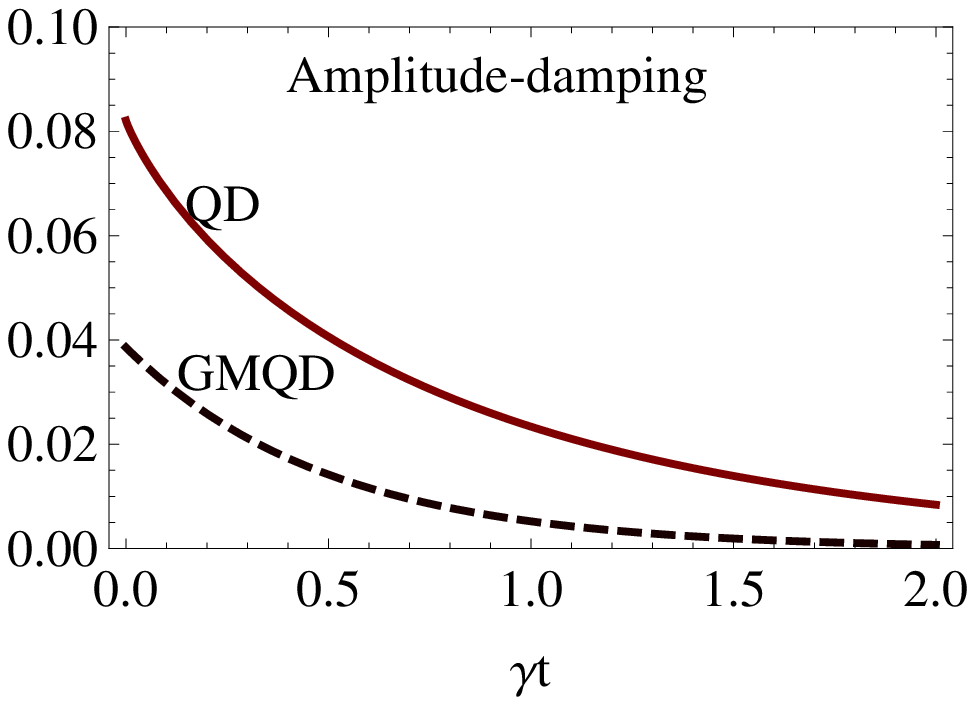}
     \includegraphics[width=7.5cm]{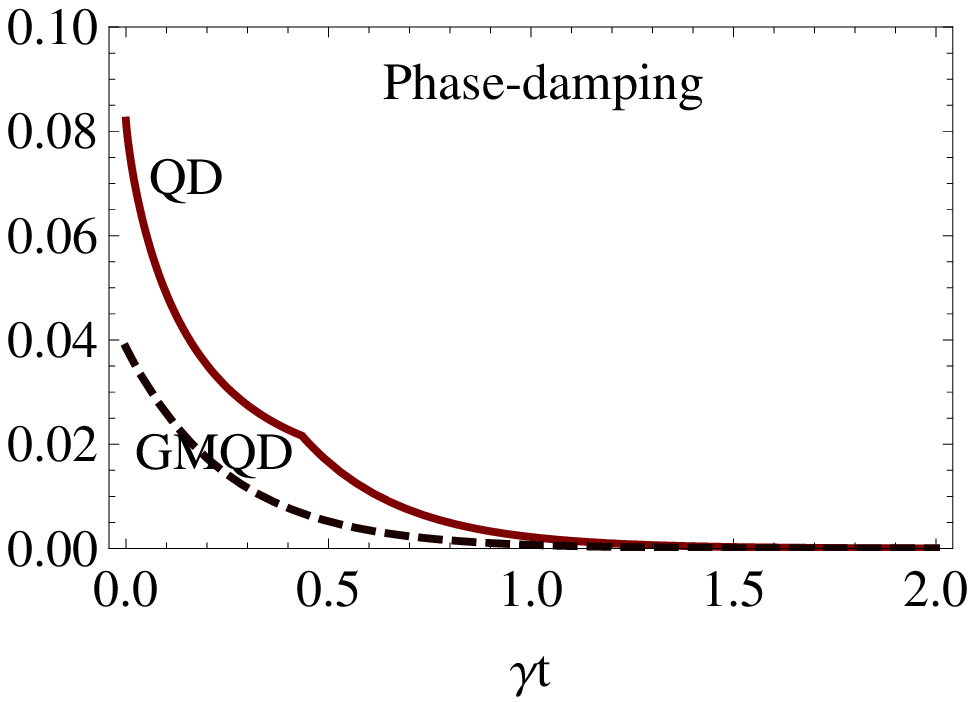}
    \includegraphics[width=7.5cm]{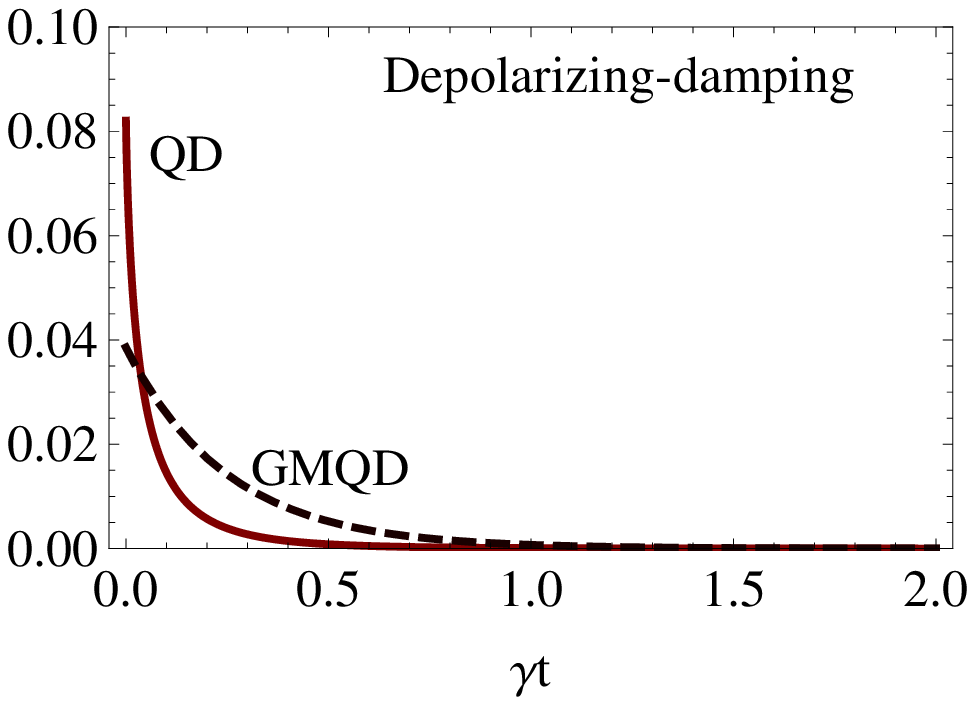}\quad
 \caption{\label{Fig3}(Color online) Dynamics of quantum discord (QD) and geometric measure of quantum discord (GMQD)
 for  the one-axis twisting collective spin state given by Eq. (11) with $\theta=0.1\pi$, $N=12$ in different types of noisy channels.}
\end{center}
\end{figure}

Next, we investigate the influence of initial angle $\theta$  of the one-axis twisting collective state on the dynamics of the pairwise QD under different types of noisy channels. We plot the pairwise QD of the one-axis twisting collective spin state given by Eq. (11) as a functions of $\gamma t$ and $\theta$ for fixed $N=12$ in different types of noisy channels, as shown in Figure 2. It is clearly seen that, in phase-flip channel, the pairwise QD quickly decays to zero and revives. When the amount of the pairwise QD arrives at its initial value, it is not effected by phase-flip noisy channel. This indicates that QD for pairs of particles extracted from a symmetric state of multi-qubit systems can persist at a steady state. Besides, the behavior of the pairwise QD is symmetrical during a period $2\pi$. However, compared to phase-flip channel, the pairwise QD dynamics decay in a monotonic fashion in amplitude damping, Phase-damping, and depolarizing channel.

Finally, we compare the pairwise QD dynamics with the geometric measure of quantum discord (GMQD) for a symmetric multi-qubit system in different types of noisy channels. Figure 3, presents dynamics of QD and GMQD
for the one-axis twisting collective spin state given by Eq. (11) with $\theta=0.1\pi$, $N=12$ in different types of noisy channels. It is also observed that the  pairwise QD has subtle differences with GMQD. Quantum correlations quantified by QD are always larger than those quantified by GMQD in decoherence noisy channels except for depolarizing noisy channel. This means the QD is more robust than GMQD under the influence of decoherence noisy channels, and depolarizing noisy channel heavily influences the pairwise quantum correlations dynamics.

\section{Conclusion}  
We study the pairwise QD for a symmetric multi-qubit system in  different types of noisy channels, such as phase-flip, amplitude damping, phase-damping, and depolarizing channels. Using QD and GMQD to quantify quantum correlations, some analytical or
numerical results are presented. The results show that, the dynamics of  the pairwise QD is related to the number of spin particles $N$ as well as initial parameter $\theta$ of the one-axis twisting collective state. With the number of spin particles $N$ increasing, the amount of the pairwise QD increases. However, when the amount of the pairwise QD arrives at a stable maximal value, it is independence of the number of spin particles $N$ increasing. The behavior of the pairwise QD is symmetrical during a period $2\pi$. Moreover, we compare the pairwise QD dynamics with the geometric measure of GMQD for a symmetric multi-qubit system in different types of noisy channels. It is perhaps helpful for the quantum information processing.

\acknowledgments
This work is supported by the National Natural Science Foundation of China (Grant Nos. 11374096 and 11074072) and Scientific Research Foundation of Hunan Provincial Education Department (No. 13C039).

\label{app:eff-trans}

\end{document}